\documentclass[conference]{IEEEtran}
\IEEEoverridecommandlockouts

\usepackage[top=0.75in, left=0.6in, right=0.7in, bottom=1.05in]{geometry}
\usepackage[nolist]{acronym}
\usepackage{amsmath}

\usepackage{mathtools}
\usepackage[linesnumbered,ruled,vlined]{algorithm2e}
\usepackage{graphics}
\usepackage{breqn}
\usepackage{amsmath, amssymb}
\usepackage{algorithmic}
\usepackage{cite}
\usepackage{graphicx}

\graphicspath{{Figures/}}
\usepackage{xcolor} 

\usepackage{subcaption}

\begin{document}

\title{Green O-RAN Operation: a Modern ML-Driven Network Energy Consumption Optimisation}

\author{\IEEEauthorblockN{Xuanyu Liang\IEEEauthorrefmark{1}, Ahmed Al-Tahmeesschi\IEEEauthorrefmark{1}, Swarna Chetty\IEEEauthorrefmark{1} and Hamed Ahmadi\IEEEauthorrefmark{1}}
\\
\IEEEauthorrefmark{1}School of Physics Engineering and Technology, University of York, United Kingdom\\
}

\maketitle

\thispagestyle{empty}
\pagestyle{empty}
\begin{abstract}
The increasing energy demand of next-generation mobile networks, especially 6G, is becoming a major concern—particularly due to the high power usage of base station components \acp{RU}, which often remain active even during low traffic periods. To tackle this challenge, our study focuses on improving energy efficiency in \ac{O-RAN} systems using intelligent control strategies. \ac{TD3} leverages a continuous action space to overcome the limitations of traditional discrete-action methods like \ac{DQN}. By avoiding exponential growth in action space, TD3 enables more precise control of RU sleep modes in dense and large radio environments. Simulation results show that our approach consistently achieves over 50\% energy savings compared to the always-on baseline, with TD3 outperforming DQN-based methods by up to 6\%, while also offering better stability and faster convergence.

\end{abstract}


\section{Introduction}

To accommodate the rapidly increasing demand for mobile traffic, \acp{BS} have been widely deployed to satisfy user data rate requirements. Although dense deployments of \acp{BS} significantly enhance network capacity and ensure better coverage and connectivity, they also lead to substantial energy consumption. Approximately 80\% of the energy consumption in mobile cellular networks is attributed to \acp{BS}. The increase in energy consumption not only amplifies operational costs for wireless network providers but also contributes to an increase in CO2 emissions \cite{liang2024energy}. Consequently, addressing energy efficiency in BS operations is becoming increasingly crucial. Traditional \ac{BS} designs often feature integrated hardware with limited flexibility, making it difficult to manage and optimize energy use effectively. Such integrated designs limit the ability to selectively activate or adjust the power levels of individual components, resulting in a continuous and frequently inefficient energy consumption. 

In this context, the \ac{O-RAN} introduces a novel architecture that disaggregated traditional \ac{BS} functionalities into three key components: the \ac{CU}, the \ac{DU}, and the \ac{RU} and the integration of open interfaces and standards \cite{O-RanArchitecture, O-RanA1Interface}. This architectural change allows for more granular control over network functionality, enabling better scalability, easier upgrades, and more targeted energy-saving measures. By hosting the CU and DU functions in the cloud and simplifying the role of the RU to focus on lower-layer physical processing and RF tasks, O-RAN enables a more flexible and resource-efficient deployment. Another key innovation in \ac{O-RAN} is the introduction of the \ac{RIC}, which addresses various operational dynamics through its two main components: the \ac{Near-RT RIC} and the \ac{Non-RT RIC}. In this paper, we focus on the development and deployment of xApps within the \ac{Near-RT RIC}. xApps are modular, third-party applications that run on the Near-RT RIC platform and enable real-time, policy-driven control of radio resources.

The energy consumption of \acp{RU} represents a significant fraction of the overall energy used by cellular networks, particularly in dense 5G deployments \cite{abubakar2023energy}. Although \acp{RU} are primarily deployed to handle peak traffic demands, their continuous operation during periods of low traffic results in substantial energy wastage. To mitigate this inefficiency, dynamically switching off lightly-loaded \acp{RU}—a strategy commonly referred to as sleep mode has emerged as an effective energy-saving approach \cite{wu2015energy,liu2015small,oh2016unified,peng2014stochastic}. For example, \cite{liu2015small} introduced a random sleep mode mechanism for \acp{BS} in small cell networks, while \cite{oh2016unified} proposed a sequential algorithm to deactivate \acp{BS} with minimal network impact while meeting mobile rate requirements. Additionally, \cite{peng2014stochastic} presented several schemes for deactivating macro \acp{BS} in heterogeneous networks, maintaining the original coverage performance by adjusting the transmission power of the remaining active macro \acp{BS} and incorporating additional micro \acp{BS}. Although mathematical optimization techniques have proven effective in enhancing network energy efficiency, their application in complex scenarios, with an increasing number of \acp{BS} and multi-objective constraints, can become computationally intensive and time-consuming.

In this context, \ac{ML} approaches have been proposed to reduce energy consumption in complex and dynamic wireless networks while preserving the \ac{UE} \ac{QoS} \cite{ye2019drag,ju2022energy,jang2020base,sun2024deep}. In particular, both \cite{jang2020base} and \cite{ye2019drag} introduce \ac{ML}-based strategies tailored for ultra-dense networks. In \cite{jang2020base}, an \ac{LSTM}-based deep neural network is employed to extract temporally correlated features from channel information, thereby enabling the dynamic switching on/off of \acp{BS}. In contrast, \cite{ju2022energy} utilizes a \ac{DQN} to optimize the sleep mode of \acp{BS} and incorporates an action selection network to reduce the action space by filtering invalid actions. Moreover, \cite{ye2019drag} proposes an actor-critic \ac{DRL} approach that addresses the challenges posed by large action spaces without relying on explicit action space reduction, while also predicting future traffic arrivals to improve decision-making accuracy.

In this paper, we propose a \ac{TD3}-based algorithm to optimize \ac{RU} sleep mode decisions for improved energy efficiency in large and dense O-RAN environments. Specifically, we design and implement the \ac{TD3} model as an xApp that operates based on real-time user activity patterns, rather than relying on static traffic snapshots as in prior work \cite{wang2024energy}. By utilizing a continuous action space, \ac{TD3} avoids the exponential growth in action combinations that occurs with traditional discrete-action methods when the number of \acp{RU} increases, enabling scalability in complex network settings. For comparative evaluation, we introduce two variants of Deep Q-Network (DQN) as baseline models: \ac{DQNMA} and \ac{DQNSA}. Simulation results demonstrate that the \ac{TD3} model achieves greater energy savings and faster convergence than both DQN-based approaches. Furthermore, unlike \ac{DQNMA}, which becomes infeasible in large-scale deployments due to action space explosion, the \ac{TD3} model remains stable and effective even in extended and complex network areas.


The main contributions of this paper are:
\begin{itemize}
    \item We develop a \ac{TD3}-based \ac{DRL} algorithm for optimizing the sleep mode strategy of \acp{RU} in O-RAN.  The proposed model is implemented as an xApp, enabling low-latency, data-driven control over RU activation states.
    \item To establish a comparative baseline, we also develop two \ac{DQN} variants: \ac{DQNMA}, which simultaneously controls all RUs, and \ac{DQNMA}, which updates one RU at each time step to reduce action space complexity.
    \item We demonstrate that the TD3 model consistently outperforms the DQN-based approaches in both energy savings and convergence speed.
\end{itemize}



\section{System Model}

\begin{figure*}[!htbp]
	\centering
	\includegraphics[clip, trim=0.0cm 0.cm 0.0cm 0cm, width=1.8\columnwidth]{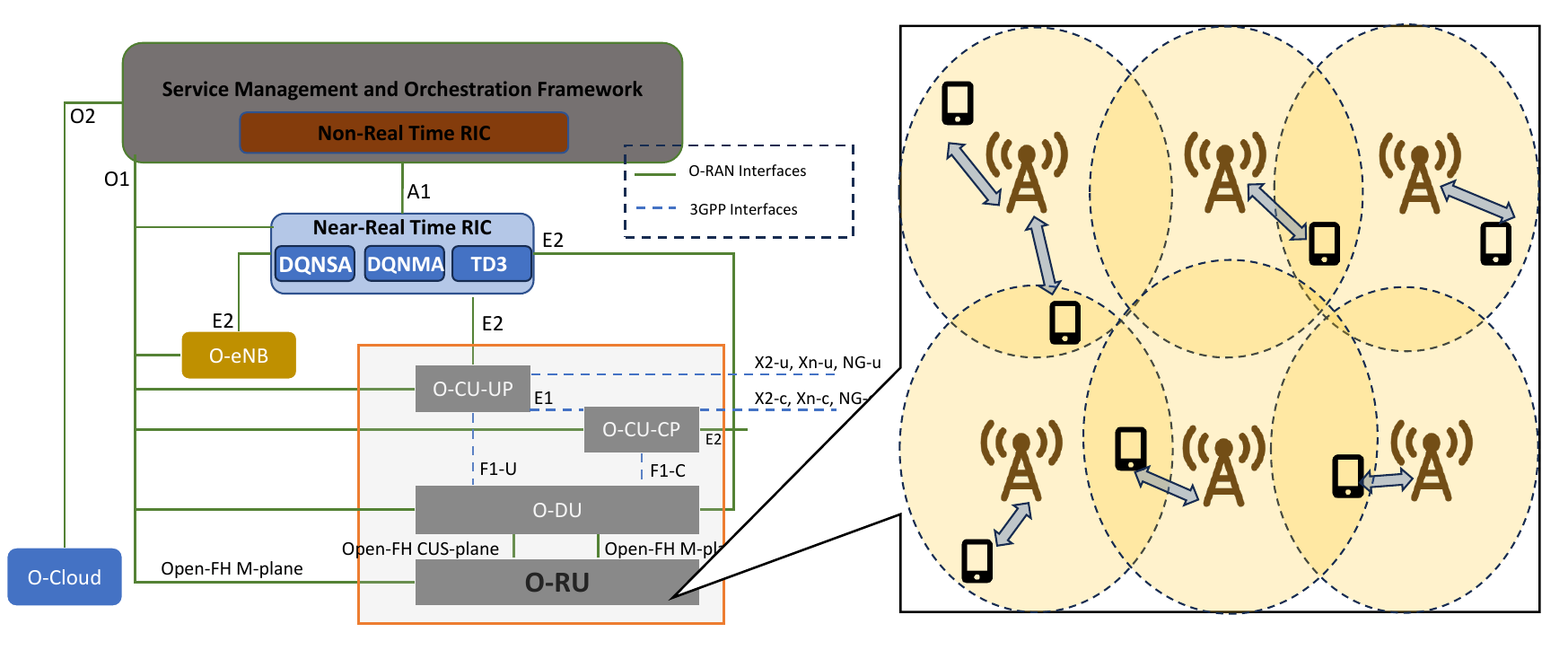}
	\caption{shows O-RAN Architecture, with components and interfaces from O-RAN and 3GPP. O-RAN interfaces are drawn as solid lines, 3GPP ones as dashed lines. The layout of the O-RU  is shown in the picture in the right section.}
	\label{fig:systemmodel}
\end{figure*}



In this section, we discuss the system model of O-RAN environment. We consider the downlink transmission where $M$ \acp{RU} equipped with a single antenna serve $K$ \acp{UE}. Each \ac{RU} is served by a corresponding \ac{DU} and \acp{DU} are connected with single \ac{CU}. In addition, \acp{UE} move at a speed $v$, choosing from $0$ to $v_{max}$.  Note that, in our study, the \acp{UE} move within the area and do not leave it. The sets of \acp{RU} and \acp{UE} are represented as $\mathcal{M} = \{ 1, 2, \ldots, M \}$ and $\mathcal{K} = \{1,2, \ldots, K\}$ respectively. We also introduce the $\alpha_m$ which is a binary variable to indicate the active/sleep of \ac{RU} $m$. Here $\alpha_m = 1$ means the \ac{RU} $m$ is in active mode, otherwise $\alpha_m = 0$.

In our work, each \ac{RU} is able to provide a total of $Q_m$ \acp{PRB} to the \acp{UE} associate to the \ac{RU}. Let $U_{m}^t$ denote the total number of \acp{UE} associate with \ac{RU} $m$ at time slot $t$. On top of that, we denote $n_{m,k}^t$ is the number of \acp{PRB} allocated to the $k^{th}$ \ac{UE} in \ac{RU} $m$. Therefore, the total number of \acp{PRB} usage $N_{m}^{t}$ of \ac{RU} $m$ at time slot $t$ can be computed as $N_{m}^{t} = \sum_{u = 0}^{U_m^t} n_{m,k}^t$. Based on this, the \ac{RU} load at time slot $t$ is defined as:
\begin{equation}\label{2}
    l_{m}^{t} = \frac{N_{m}^{t}}{Q_m}.
\end{equation}
In our work, we consider the fading channel model from the \ac{RU} $m$ to the \ac{UE} $k$ as $h_{m,k}$. Therefore the \ac{SNR} $SNR_{m,k}$ of the $k^{th}$ \ac{UE} associate to the \ac{RU} $m$ at time $t$ is presented as :
\vspace{-2mm} 
\begin{equation}
    SNR_{m,k} = \frac{P_{TX}h_{m,k}\frac{n_{m,k}}{Q_m}}{n_{m,k}N_0},
\end{equation}

where $P_{TX}$ is the maximum transmission power of an \ac{RU} and the transmission power allocated to the \ac{UE} depends on the percentage of \acp{PRB} occupancy. The denominator corresponding to additive white Gaussian noise. The corresponding data rate of \ac{UE} $k$ is given by:
\vspace{-2mm}  
\begin{equation}
    R_{k}^{t} = n_{m,k}B * \log_2(1 + SNR_{m,k}),
\end{equation}

where $B$ is the bandwidth of each \ac{PRB}

\subsection{Power Consumption Model and Problem Formulation}
The power consumption of \ac{RU} is divided into three parts. The fisrt part is the fixed power $P_{m}^{Fix}$ of \acp{RU} consumed by the signal processing, cooling system and power supply, which only varies depending on the active/sleep mode of \ac{RU}. Then the load dependent power comes from the \ac{PA} $P_{m}^{data}$. The load in this paper is represented by the \ac{PRB} usage as shown in \eqref{2}, and transition power $P_{m}^{trans}$ is generated when the mode of the base station changes, e.g. from sleep mode to active mode. Combining these three parts, the power consumption of the \ac{RU} $m$ at time slot $t$ is:
\vspace{-1mm}  
\begin{equation}
    P_{m}^{t} = P_{m}^{Fix,t} + P_{m}^{data,t} + P_{m}^{trans,t}.
\end{equation}
\vspace{-2mm}  

First, fixed power of \ac{RU} $m$ consumed by the active or sleep mode is given by:
\begin{equation}
    P_{m}^{Fix,t} = \alpha_m^t P_m^{active} + (1 - \alpha_m^t) P_m^{sleep},
\end{equation}
where $P_m^{active}$ and $P_m^{sleep}$ are represented the fixed power consumption of \ac{RU} $m$ in active mode and sleep mode, respectively.
Then, the transmission power of \ac{RU} $m$ which is scaled with load of \ac{RU} $m$ at time slot $t$ computed as:
\begin{equation}
    P_{m}^{data,t} = \alpha_m^t \frac{P_{TX}}{\eta} * l_m^t = \alpha_m^t \frac{P_{TX}}{\eta} * \frac{N_{m}^t}{Q_m},
\end{equation}
where $\eta \in [0,1] $ is the power amplifier efficiency of the \ac{RU} $m$. $P_{TX}$ here is the constant value and only scale with the percentage of the load. Therefore, the more \acp{UE} are associated the more power consume.
Last but not least, the transition power of \ac{RU} $m$ is given by: 
\begin{equation}
    P_{m}^{trans,t} = \max\left( \alpha_m^t - \alpha_m^{t-1}, 0 \right) \cdot V_m^{trans}
\end{equation}

where, $\alpha_m^{t-1}$ is the active/sleep mode of \ac{RU} $m$ in the last time slot. $V_m^{trans}$ is the power consumed by switching on and off mode of the \ac{RU} $m$. The transition power $P_m^{trans}$ is only consumed when the \ac{RU} is switched on.

Finally, the total power consumption of the entire network in time slot $t$ is defined as:

\vspace{-2mm}  

\begin{equation}
\begin{aligned}
P_{\text{tot}}^t & = \sum_{m=1}^{M}(P_m^{Fix,t} + P_m^{data,t} + P_{m}^{trans,t})\\
&= \sum_{m=1}^{M}(\alpha_m^t P_m^{active} + (1 - \alpha_m^t) P_m^{sleep}) + \alpha_m^t \frac{P_{TX}}{\eta} * \frac{N_{m}^t}{Q_m} \\
& \quad + ( \max\left( \alpha_m^t - \alpha_m^{t-1}, 0 \right) \cdot V_m^{trans})
\end{aligned}
\end{equation}
\vspace{-2mm}

We assume all the \acp{RU} are in active at the initial stage when $t-1 = 0$. Therefore, the power minimization problem can be formulated as follow:
 
\vspace{-2mm}  
\begin{equation}
\label{objective function}
\begin{aligned}
\mathcal{P}_1: \quad \min_{\{\alpha_{m}^{t}, n_{m,k}^{t}\}} \quad \sum_{t=1}^{T} & P_{tot}^t \\
\text{s.t.} \quad &  R_{k}^t\geq R_{k,min}^t, \quad \forall k \in \mathcal{K}, \forall t \in \mathcal{T} \\
                    & N_m^t\leq Q_m^t, \quad \forall m \in \mathcal{M}, \forall t \in \mathcal{T}\\
                    & \alpha_m^t \in \left \{ 0,1 \right \}, \quad m \in \mathcal{M}, \forall t \in \mathcal{T}, 
\end{aligned}
\end{equation}
where $R_{k,min}^t$ is the minimum data rate requirement of \ac{UE} $k$ at time interval $t$. The second constraint is that the allocated \acp{PRB} need to be within the total number of \acp{PRB} and $\mathcal{T} = \{ 1, 2, \ldots, T \}$.



\section{Energy Efficiency \ac{DRL} based Solution}

The primary objective of this study is to optimize the sleep and active modes of \ac{BS} in dynamic radio environments to minimize network-wide energy consumption without impacting the \acp{UE} \ac{QoS}. In this section, we proposed a solution combined with \ac{TD3} algorithm, an advanced actor-critical \ac{RL} framework to optimize the sleep mode strategy.

\subsection{Markov Decision Process Problem}
To solve the sleep mode optimization problem which is the NP-hard problem, we decompose the original problem into \ac{MDP} problem. In this subsection, we discuss the state space, action space, and reward function of TD3 model to consist the \ac{MDP} problem.

\textit{State Space}: state comprises essential information used for policy training. At each time step \( t \), it contains the real data rate of the \ac{UE}, represented as $
\mathbf{R}^{t} = \begin{bmatrix} R_{1}^{t}, R_{2}^{t}, \dots, R_{K}^{t} \end{bmatrix}^{T}
$,
which captures the system's ability to meet user demands. The \ac{RU} operation mode from the previous time step are denoted as
$
\boldsymbol{\alpha}^{t-1} = \begin{bmatrix} \alpha_{1}^{t-1}, \alpha_{2}^{t-1}, \dots, \alpha_{M}^{t-1} \end{bmatrix}^{T}$. The \ac{PRB} utilization of the RUs is given by
$\mathbf{L}^{t} = \begin{bmatrix} l_{1}^{t}, l_{2}^{t}, \dots, l_{M}^{t} \end{bmatrix}^{T}$,
quantifying the load on the RUs. Finally, the spatial positions of the UEs in the network are represented as:
$\mathbf{U}^{t} = \begin{bmatrix} (x_{1}^{t}, y_{1}^{t}), (x_{2}^{t}, y_{2}^{t}), \dots, (x_{K}^{t}, y_{K}^{t}) \end{bmatrix}^{T}$,
characterized by their \( (x, y) \) coordinates, providing insight into \ac{UE} distribution and mobility. In summary, the state can be expressed as:
\vspace{-2mm}  
\[
s_t = \begin{bmatrix} \mathbf{R}^{t}, \boldsymbol{\alpha}^{t-1}, \mathbf{L}^{t}, \mathbf{U}^{t} \end{bmatrix}^{T}.\]
All state parameters are normalized to facilitate a better interpretation by the agent.

\textit{Action Space}: \ac{TD3} addresses a fundamental limitation of conventional discrete-action methods such as \ac{DQN}. In such methods, the size of the action space increases exponentially with the number of \acp{BS}, scaling as $2^M$. For instance, in a system with 12 \acp{RU}, there would be $2^{12}$ potential sleep/active state combinations, making exploration computationally infeasible. In contrast, \ac{TD3} employs a continuous action space, which scales linearly with the number of \acp{RU}. So in TD3, action \( \alpha \in \mathcal{A} \) represents the binary operational states of all RUs, defined as:
\[
\mathcal{A} = \left[\alpha_1, \alpha_2, \dots, \alpha_{M}\right].
\]
Each element \( \alpha_m \) corresponds to an RU, where \( \alpha_m = 1 \) denotes that the RU is active, and \( \alpha_m = 0 \) indicates that the RU is in sleep mode. These actions are derived from the output of the \ac{TD3} model, which guides the energy-efficient operation of the network.

\textit{Reward Function }: reward \( r \in \mathbb{R} \) is designed to balance energy efficiency and user satisfaction, guiding the model towards an optimal operational policy. It is defined as:
\[
r = - w_1 \cdot \frac{P_{\text{tot}}}{P_{\text{max}}} - w_2 \cdot \frac{K_{\text{unsat}}}{K},
\]
where and \( P_{\text{max}} \) corresponds to the energy consumption if all \acp{RU} are all active. The term \( K_{\text{unsat}} \) denotes the number of users with data rates below their required thresholds, normalized by the total number of users \( K \). The weights \( w_1 \) and \( w_2 \) adjust the relative importance of energy efficiency and user satisfaction. Normalization is used in the reward function to ensure that the contributions of energy consumption and user satisfaction are scaled to comparable ranges.

\subsection{Energy efficiency TD3 solution in O-RAN}

In this work, we utilize a customized version of \ac{TD3} to dynamically control the sleep/active states of \acp{RU} in an O-RAN network, aiming to minimize energy consumption while ensuring \ac{UE} \ac{QoS}. Unlike generic applications of TD3, our approach explicitly encodes the physical behavior of base station sleep transitions and the spatio-temporal dynamics of user distribution into the RL framework as shown in Algorithm~\ref{alg:td3}.

In our problem, the action space corresponds to a binary vector representing the states of all \acp{RU}. The TD3 actor network outputs a continuous vector \( \mu_{\theta}(s) \in [0,1]^n \). This continuous output is then discretized by a thresholding function \( f_d \), such that the final action vector is:
\[
a = f_d(\mu_{\theta}(s) + \eta_{n}), \quad a_i = \begin{cases}
1 & \text{if } \mu_{\theta}(s)_i + \eta_i > 0.5 \\
0 & \text{otherwise}
\end{cases}
\]
where \( \eta_{n} \sim \mathcal{N}(0, \sigma^2) \) is temporally-decayed Gaussian exploration noise. This design allows the actor to learn a \textit{soft preference} over base station activation while the environment enforces hard switching decisions.

The environment is modeled to reflect practical conditions, including dynamic user mobility, soft handovers, and energy consumption computation. Each transition \( (s, a, r, s') \) is collected into a replay buffer. To mitigate overestimation bias, TD3 employs twin Q-networks \( Q_{\phi_1}, Q_{\phi_2} \). In our setting, this is particularly important because suboptimal base station activations may seem energy-efficient but degrade user satisfaction. The critic update targets are computed as: $y = r + \gamma \min_{i=1,2} Q_{\phi_i}(s', f_d(\mu_{\theta'}(s') + \epsilon))$
where \( \epsilon \) is clipped noise added for \textit{target policy smoothing}, encouraging robustness to small perturbations in base station decisions. This is crucial in our environment due to the discrete switching and spatial sensitivity of user association.

The actor network is updated with the deterministic policy gradient using the critic feedback, $\nabla_{\theta} J(\theta) = \mathbb{E}_{s \sim \mathcal{D}}[\nabla_a Q_{\phi}(s, a) \big|_{a = \mu_{\theta}(s)} \cdot \nabla_{\theta} \mu_{\theta}(s)]$. To improve training stability, we follow the \ac{TD3} principle of delayed policy updates, updating the actor and target networks only every steps, allowing the critics to stabilize before policy updates influence the network topology.
We also implement \textit{soft updates} for the target networks: $\theta' \leftarrow \tau \theta + (1-\tau)\theta', \quad \phi_i' \leftarrow \tau \phi_i + (1-\tau)\phi_i'$.
This smooth evolution of target networks is key in our simulation-based setup, where abrupt Q-value changes may lead to erratic base station switching.

\begin{algorithm}[t]\small
\caption{TD3-Based Energy-Aware RU Control}
\label{alg:td3}
\KwIn{Initial network state $s_0$, actor parameters $\theta$, critic parameters $\phi_1, \phi_2$, replay buffer $\mathcal{D}$}
\KwOut{Trained actor network $\mu_\theta$ for RU sleep scheduling}
\For{each episode}{
    Initialize environment and receive initial state $s$ \;
    \For{each time step $t = 1, \dots, T$}{
        Add exploration noise $\eta_t \sim \mathcal{N}(0, \sigma^2)$ \;
        Compute continuous action: $a_c = \mu_{\theta}(s) + \eta_t$ \;
        Discretize: $a = f_d(a_c)$ where $a_i = \mathbb{I}[a_{c,i} > 0.5]$ \;
        Execute action $a$ in the environment \;
        Receive reward $r$ and next state $s'$ \;
        Store $(s, a, r, s')$ into replay buffer $\mathcal{D}$ \;
        Sample a mini-batch $(s_j, a_j, r_j, s_j')$ from $\mathcal{D}$ \;
        Compute target action: $\tilde{a}_j' = f_d(\mu_{\theta'}(s_j') + \epsilon)$ \;
        Compute target Q-value: $y_j = r_j + \gamma \min_{i=1,2} Q_{\phi_i'}(s_j', \tilde{a}_j')$ \;
        Update critics $\phi_1, \phi_2$ by minimizing:
        $\mathcal{L}(\phi_i) = \frac{1}{N} \sum_j \left( Q_{\phi_i}(s_j, a_j) - y_j \right)^2$ \;
        \If{every $d$ steps}{
            Update actor using deterministic policy gradient \;
            Soft-update target networks: $\phi_i' \leftarrow \tau \phi_i + (1-\tau)\phi_i'$, $\theta' \leftarrow \tau \theta + (1-\tau)\theta'$ \;
        }
        $s \leftarrow s'$
    }
}
\end{algorithm}

\section{Simulation Results}

\begin{figure}[!t]
	\centering
	\includegraphics[clip, trim=0.0cm 0.cm 0.0cm 0cm, width=0.8\columnwidth]{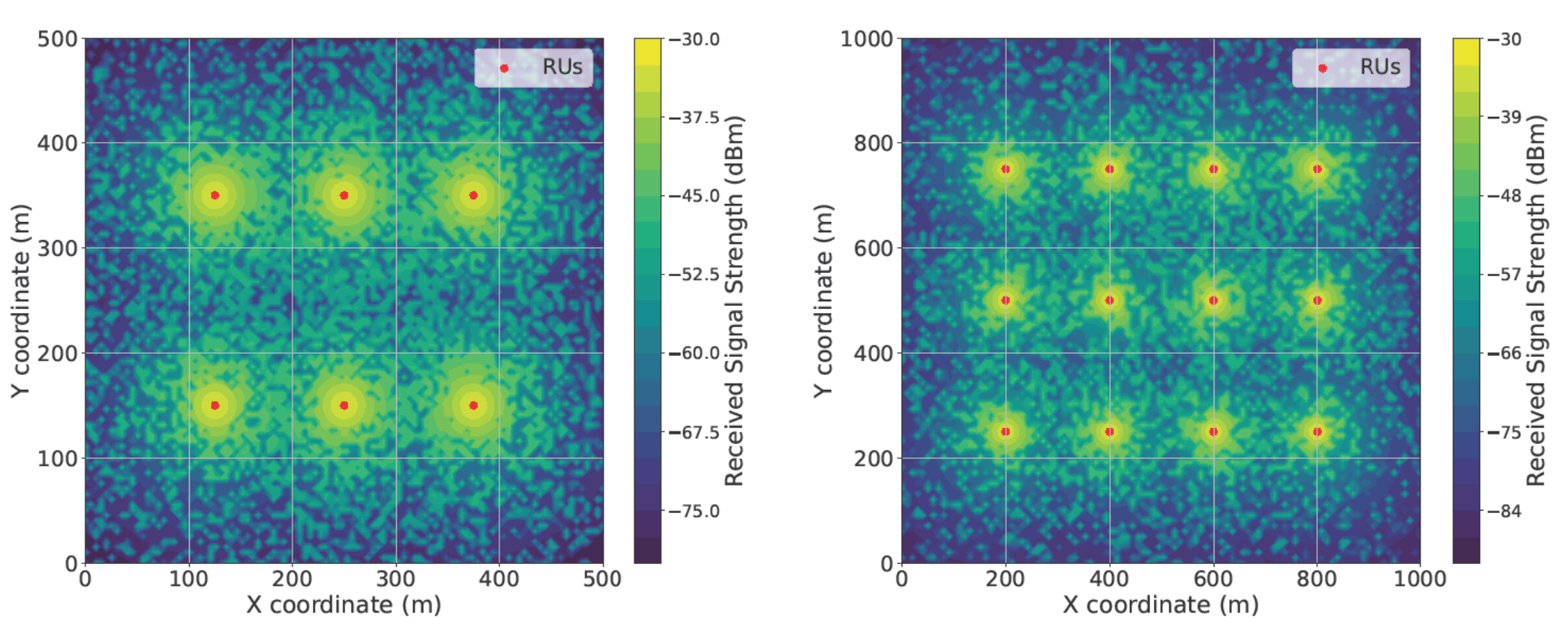}
	\caption{illustrates the radio maps of 500 $\times$ 500 m$^2$ and 1000$\times$ 1000 m$^2$ area.}
    \label{fig:2scenario-radiomap}
\end{figure}

\subsection{Simulation Settings}

In this section, we present numerical results to evaluate the performance of the proposed TD3-based model. Our simulations are conducted within an \ac{O-RAN} environment, where $M$ \acp{RU} serve $K$ \acp{UE}. The \acp{RU} are uniformly spaced across a square service area of dimensions $L \times L$ m$^2$ illustrated in Fig.\ref{fig:2scenario-radiomap}. To investigate the impact of network size on model performance, we consider two different service area sizes. This comparison allows us to assess how performance varies in different spatial environments. It is important to note that the DQN multi-action model is constrained to small service areas due to the exponentially growing action space in larger environments. Consequently, its training feasibility is limited when the network size increases, whereas the other models remain applicable across different service area scales. The \acp{UE} move at a speed, denoted as $v = v_{\text{avg}} \pm v_{\text{std}}$, where $v_{\text{avg}}$ represents the mean speed and $v_{\text{std}}$ denotes the standard deviation of the speed. The actual speed of each \ac{UE} is randomly selected within this range. Furthermore, we define a movement probability distribution for \acp{UE} based on their location to ensure a periodic mobility pattern. This pattern dictates that \acp{UE} consistently move from the edge of the service area toward the center and then back to the edge, maintaining a structured and cyclic movement behavior.
For the fading channel model, we employ the \ac{UMi} channel model, considering the conditions \ac{LOS} and \ac{NLOS}, expressed as:

For LOS conditions:
\begin{equation}
\begin{array}{l}
PL_{m,k}^{\text{LOS}} = \\
\begin{cases}
32.4 + 21 \log_{10}(d_{m,k}) + 20 \log_{10}(f), & d_{m,k} \leq d_{\text{BP}}, \\
32.4 + 40 \log_{10}(d_{m,k}) + 20 \log_{10}(f) - 9.5, & d_{m,k} > d_{\text{BP}},
\end{cases}
\end{array}
\end{equation}

\begin{equation}
    d_{\text{BP}} = \frac{4 h_{\text{RU}} h_{\text{UE}} f}{c},
\end{equation}
where $d_{\text{BP}}$ is the breakpoint distance, $PL_{m,k}$ represents the path loss, which follows the UMi path loss model.

For NLOS conditions:
\begin{equation}
\begin{split}
    PL_{m,k}^{\text{NLOS}} = &\,35.3 + 22.4 \log_{10}(d_{m,k}) \\
    & + 21.3 \log_{10}(f) - 0.3(h_{\text{UE}} - 1.5),
\end{split}
\end{equation}
where $d_{m,k}$ denotes the distance between the RU $m$ and the UE $k$, $f$ is the carrier frequency in GHz, and $h_{\text{UE}}$ is the height of the UE in meters.

The LOS probability is determined by:
\begin{equation}
    P_{\text{LOS}} = \min \left( \frac{18}{d_{m,k}}, 1 \right) \left(1 - e^{-d_{m,k}/36} \right) + e^{-d_{m,k}/36}.
\end{equation}
The NLOS probability is then given by $P_{\text{NLOS}} = 1 - P_{\text{LOS}}$.

\begin{table}[h] \small
\centering
\resizebox{\linewidth}{!}{ 
\begin{tabular}{|c|c|}
\hline
\textbf{Parameters}               & \textbf{Value}  \\ \hline
Carrier frequency ($f$)              & 2 GHz           \\ \hline
RU height ($h_{RU}$)                       & 15 m            \\ \hline
UE height ($h_{UE}$)                       & 1.7 m           \\ \hline
Network size ($L$)                   & {[}500,1000{]} m \\ \hline
Number of RUs ($M$)                   & {[}6, 12{]}     \\ \hline
Number of UEs ($K$)                   & {[}10 - 80{]}   \\ \hline
Minimum Data rate requirement ($R_{\text{min}}$)                & 3 Mbps            \\ \hline
Noise power ($\sigma_{n}^{2}$)                     & -174 dBm/Hz     \\ \hline
Power amplifier ($\eta$)                & 0.5             \\ \hline
Average speed of UE ($v_{\text{avg}}$)             & 2 m/s           \\ \hline
Standard deviation of UE speed ($v_{\text{std}}$) & 0.5 m/s         \\ \hline
Fix active mode RU power($P^{\text{active}}$)        & 20 W            \\ \hline
Fix sleep mode RU power ($P^{\text{sleep}}$)         & 5 W             \\ \hline
Maximum transmission power ($P_{\text{TX}}$)      & 1 W / 30dBm     \\ \hline
Mode transition power ($P^{trans}$)           & 3 W            \\ \hline
\end{tabular}
}
\caption{Simulation Parameters}
\label{tab:network_architecture}
\end{table}

During the training process, each episode consists of 200 time steps, with each time step corresponding to 1 second. At every time step, the agent determines the sleep mode status of the \acp{RU}, making operational decisions accordingly. The network energy consumption is then evaluated based on these decisions. As a result, the total simulation duration for each episode amounts to 200 seconds. 
\begin{table}[!t]
\centering
\resizebox{\linewidth}{!}{ 
\begin{tabular}{|l|l|l|l|l|l|}
\hline
                & \textbf{Layer1} & \textbf{Layer2} & \textbf{Layer3} & \textbf{Layer4} & \textbf{Output Layer} \\ \hline
\textbf{Actor}  & BN+relu 512     & relu 256        & relu 128        & None            & sigmoid $M$             \\ \hline
\textbf{Critic} & relu 512        & relu 256        & relu 128        & None            & linear 1              \\ \hline
\textbf{DQNSA}  & relu 512        & relu 384        & relu 256        & relu 128        & linear $M^2$             \\ \hline
\textbf{DQNMA}  & relu 512        & relu 384        & relu 256        & relu 128        & linear $2M$             \\ \hline
\end{tabular}
}
\caption{Network Configurations}
\label{tab:Network Configuraions}
\end{table}
In the \ac{TD3} model, both the actor and critic networks consist of four fully connected layers. The configuration of activation functions and layer sizes are detailed in Table~\ref{tab:network_architecture}. Additionally, \ac{BN} is incorporated to enhance training stability. For the DQN models, both networks comprise five fully connected layers, with their detailed architectures also provided in Table~\ref{tab:Network Configuraions}.

In the subsequent section, we present a comparative analysis of the proposed TD3 model against three baseline techniques for \ac{RU} sleep mode management. The first baseline, termed "All Active Mode," serves as a reference scenario where all \acp{RU} remain continuously active. The second baseline is the \ac{DQNSA} model, which dynamically switches the sleep mode status of only one RU per time slot. The third baseline, referred to as the \ac{DQNMA} model, simultaneously adjusts the sleep modes for all \acp{RU} within the network at each time interval.


\subsection{Simulation Results}

\begin{figure}[!t]
	\centering
	\includegraphics[clip, trim=0.0cm 0.cm 0.0cm 0cm, width=0.8\columnwidth]{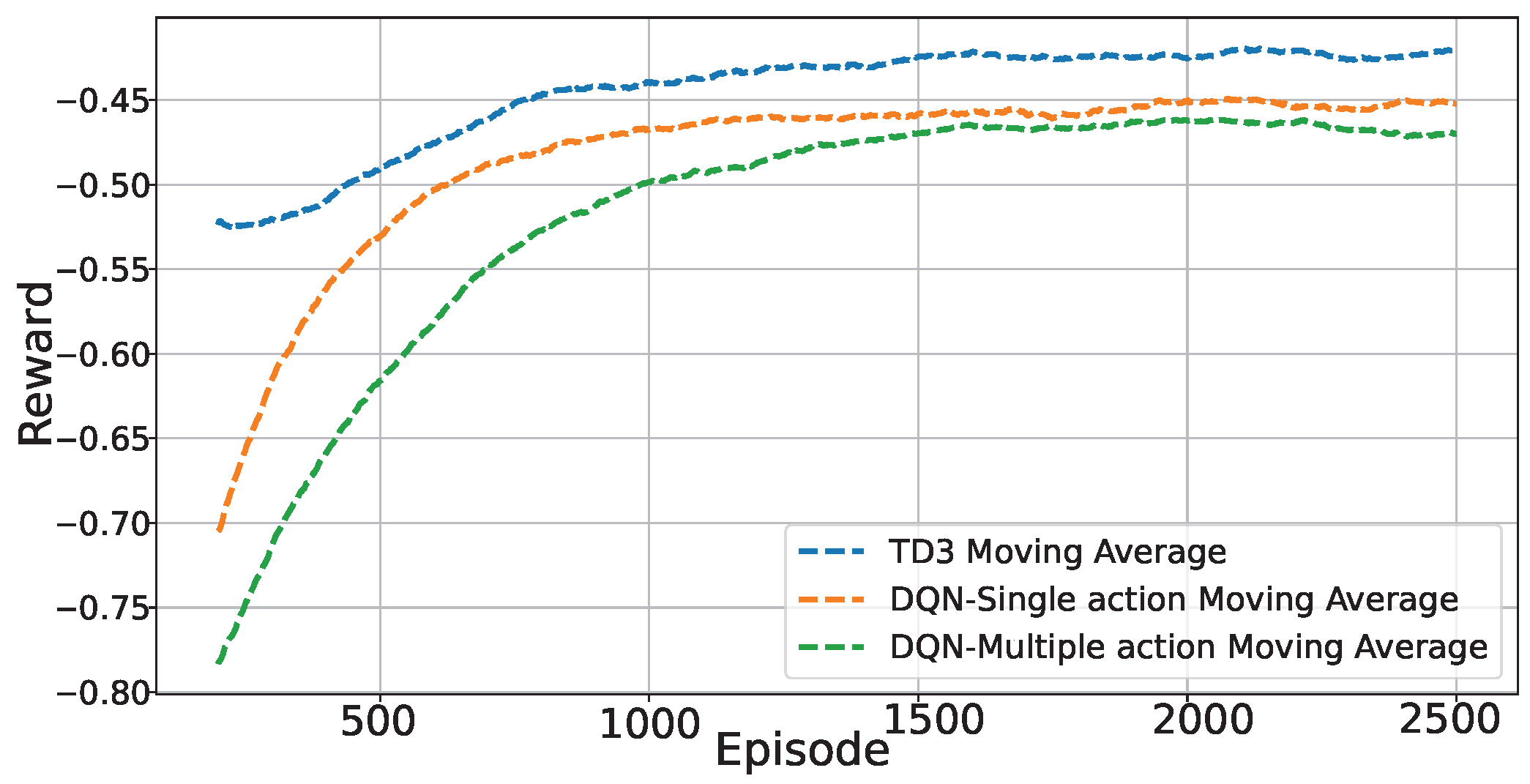}
	\caption{illustrates the rewards of TD3, DQNSA and DQNMA in 500 $\times$ 500 m$^2$ scenario}
    \label{fig:3models-500-reward}
\end{figure}

In Fig.\ref{fig:3models-500-reward} illustrates the reward performance over 2500 episodes for the \ac{TD3}, \ac{DQNSA} and \ac{DQNMA} models in 500 $\times$ 500 m$^2$ area. Notably, the TD3 model consistently achieves the highest reward performance, demonstrating superior stability and convergence compare to both DQN models. The \ac{DQNSA} model exhibits moderate performance with noticeable fluctuations, while the \ac{DQNMA} model shows the lowest reward, indicating that increasing the complexity of actions within the DQN architecture does not necessarily translate into better reward outcomes in this scenario.
\begin{figure}[!t]
	\centering
	\includegraphics[clip, trim=0.0cm 0.cm 0.0cm 0cm, width=0.8\columnwidth]{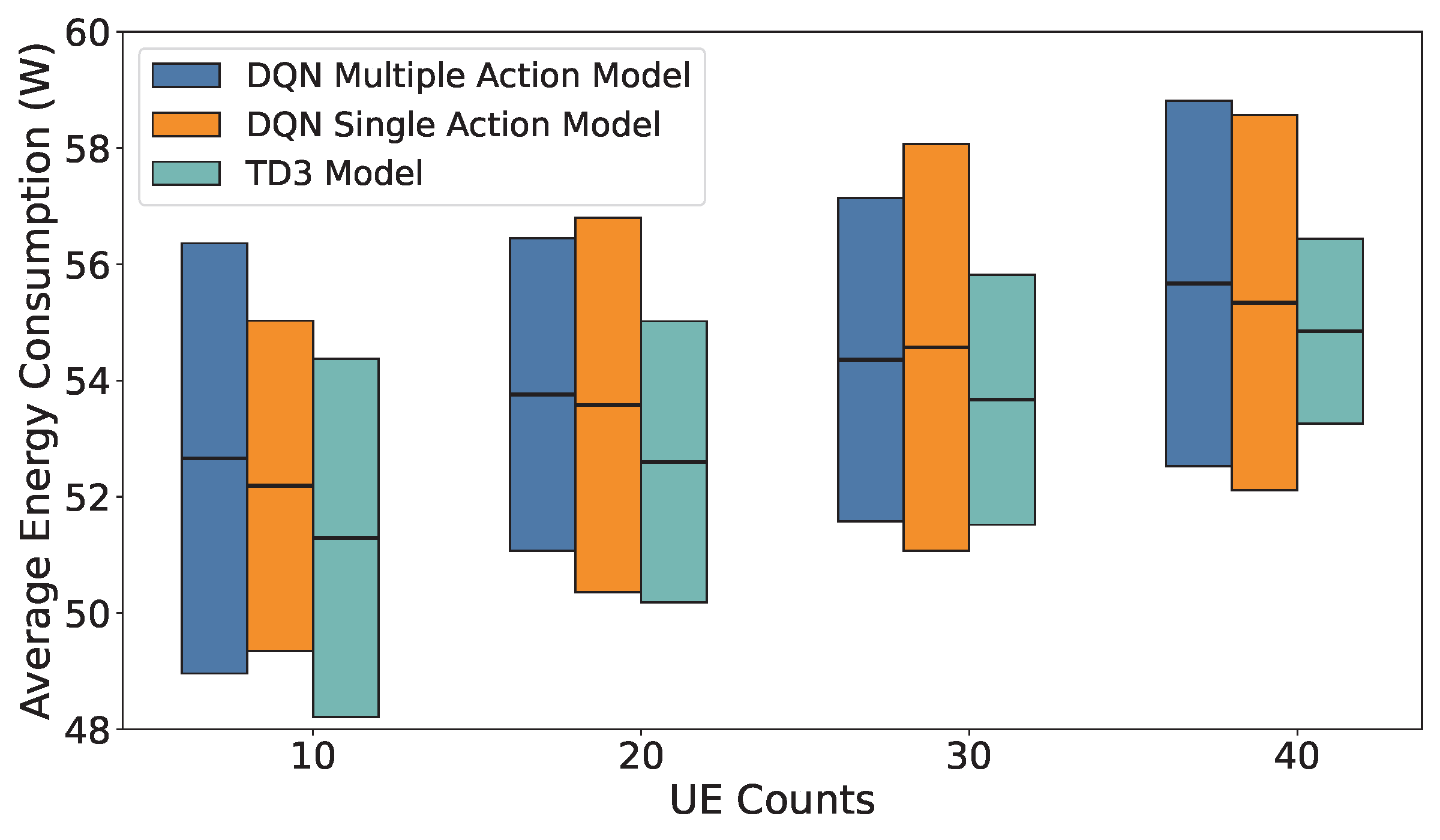}
	\caption{illustrates the average energy consumption in 500 $\times$ 500 m$^2$ scenario among 10 to 40 \acp{UE} with DQNMA, DQNSA and TD3 models.}
    \label{fig:3models-500-power}
\end{figure}
In Fig.\ref{fig:3models-500-power} indicates the energy consumption performance in 500 $\times$ 500 m$^2$ area for varying \ac{UE} counts ranging from 10 to 40, is still compared among the previous three models. The figure distinctly shows the average energy consumption, indicated by the black line in the center of each bar, while the length of each bar represents the variance in energy consumption. With a total of 6 \acp{RU} installed in this area, the maximum possible energy consumption reaches 126w when all \acp{RU} operate in active mode. The results clearly demonstrate that all three models achieve significant energy savings, exceeding 50\% compared to the theoretical maximum. Specifically, the \ac{TD3} model achieving up to an additional 6\% energy saving compared to the two \ac{DQN}-based models. Conversely, the \ac{DQNMA} model demonstrates the highest energy consumption, underscoring potential inefficiencies associated with more complex action spaces.
\begin{figure}[!t]
	\centering
	\includegraphics[clip, trim=0.0cm 0.cm 0.0cm 0cm, width=0.8\columnwidth]{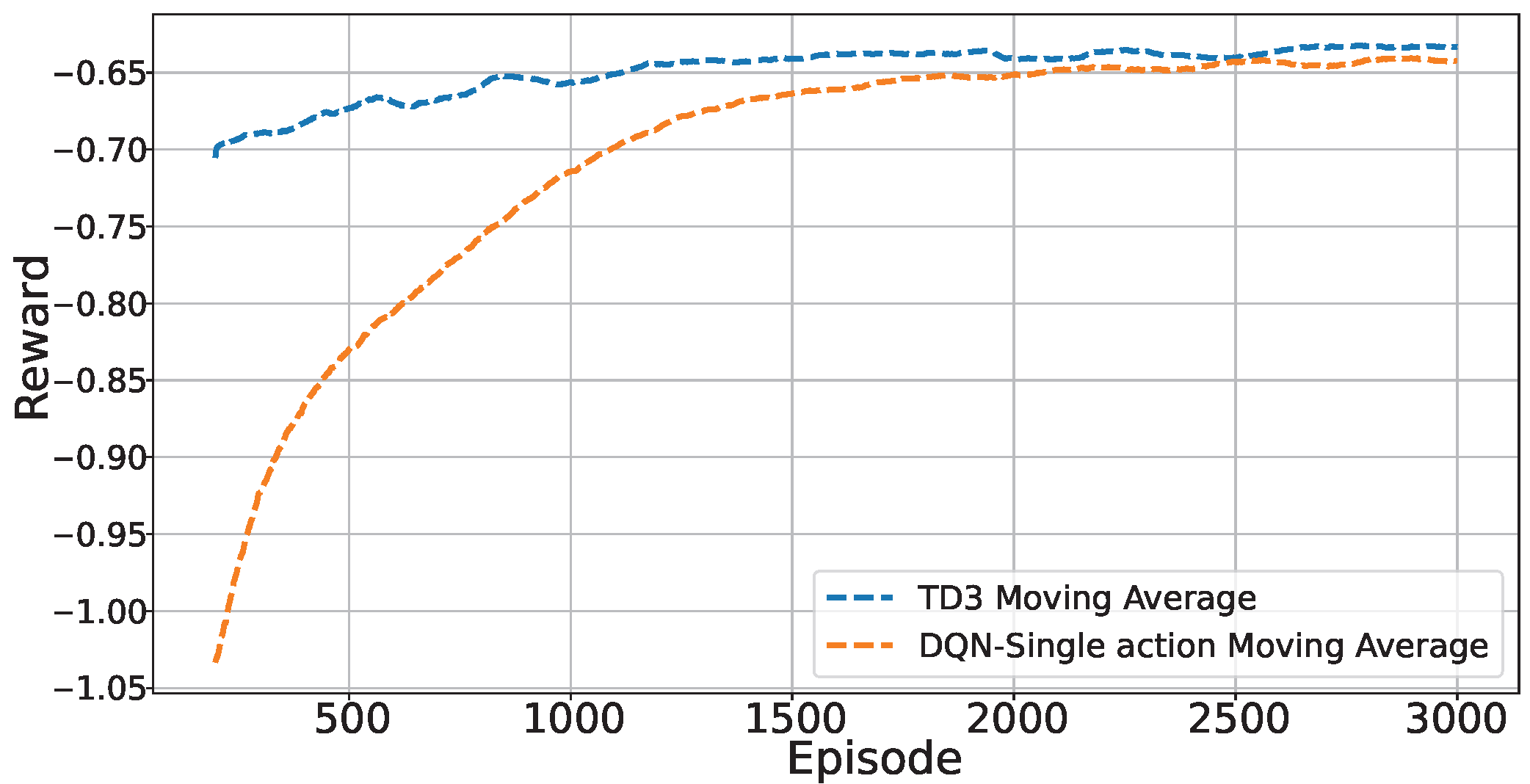}
	\caption{illustrates the rewards of TD3 and DQNSA model in 1000 $\times$ 1000 m$^2$ scenario}
    \label{fig:2models-1000-reward}
\end{figure}

\begin{figure}[!t]
	\centering
	\includegraphics[clip, trim=0.0cm 0.cm 0.0cm 0cm, width=0.8\columnwidth]{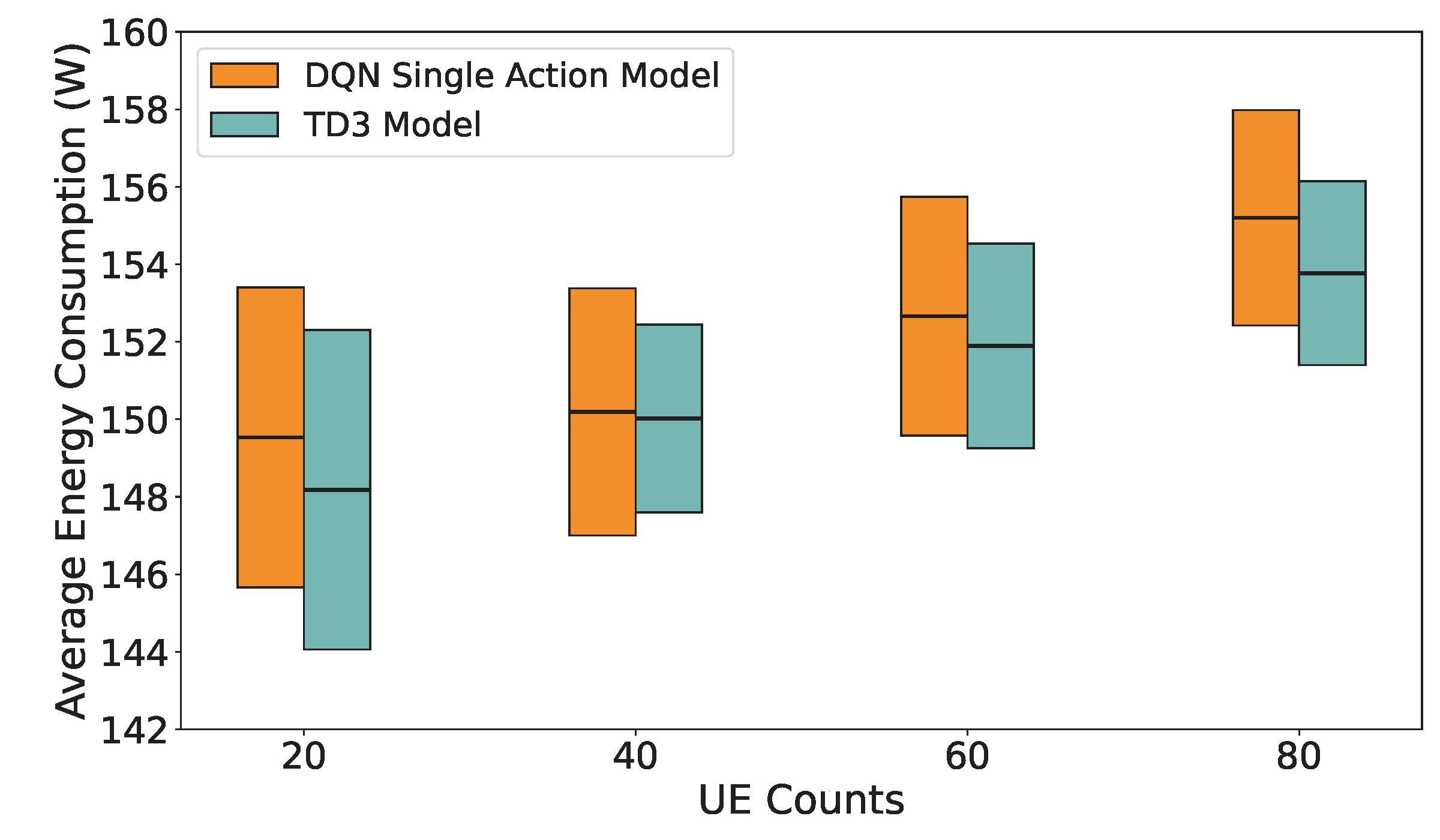}
	\caption{illustrates the average energy consumption among 20 to 80 \acp{UE} in 1000 $\times$ 1000 m$^2$ scenario with TD3 and DQNSA models.}
    \label{fig:2models-1000-power}
\end{figure}

In Fig.\ref{fig:2models-1000-reward} shows the reward performance over an extended simulation environment from 500 $\times$ 500 m$^2$ area to 1000m$^2$ area. But only compare the \ac{TD3} model against the \ac{DQNSA} model due to the excessive expansion of \ac{DQNMA} action space (e.g. 2$^{12}$). In the results, the \ac{TD3} model still achieves higher and stable reward levels throughout all episodes. Furthermore, the TD3 model demonstrates faster and more efficient convergence toward optimal solutions. The Fig.\ref{fig:2models-1000-power} illustrates the energy consumption between the \ac{TD3} and \ac{DQNSA} model, covering \ac{UE} counts from 20 to 80. Numerically, the \ac{TD3} model maintains lower energy consumption, approximately 144w at 20 UEs and up to around 152w at 80 UEs, significantly below the theoretical maximum consumption of 252 watts (more than 40\%) with all 12 RUs active. However, the TD3 model can also achieve 5\% over the \ac{DQNSA} mode, further emphasizing the TD3 model in a complex and large-scale environment.

\vspace{-2mm}
\section{Conclusions}

This paper proposed an energy-efficient framework for controlling RU sleep modes in \ac{O-RAN} using advanced \ac{DRL} methods. Proposed \ac{TD3} algorithm was implemented to effectively handle the continuous action space inherent to RU state decisions, directly addressing the limitations encountered by discrete-action methods like \ac{DQN}-based approaches. In comparison, the proposed \ac{TD3} algorithm demonstrated superior performance, consistently achieving higher stability, faster convergence, and greater energy efficiency. Numerical results clearly indicated that TD3 delivered up to 6\% additional energy savings over the baseline DQN methods, particularly highlighting its scalability and effectiveness in larger and more dynamic network scenarios. Future research directions will focus on enhancing this TD3-based approach through \ac{FL}. This future work aims to leverage \ac{FL} principles to further enhance scalability.
\vspace{-2mm}

\section{Acknowledgment}

This work has been supported by CHEDDAR: Communications Hub for Empowering Distributed Cloud Computing Applications and Research, funded by the UK EPSRC under grant numbers EP/Y037421/1 and EP/X040518/1, and by the Department of Science, Innovation and Technology, United Kingdom, under Grant Yorkshire Open-RAN (YORAN) TS/X013758/1.

\vspace{-2mm}
\begin{acronym} 
\acro{5G}{Fifth Generation}
\acro{ACO}{Ant Colony Optimization}
\acro{ANN}{Artificial Neural Network}
\acro{BB}{Base Band}
\acro{BBU}{Base Band Unit}
\acro{BER}{Bit Error Rate}
\acro{BS}{Base Station}
\acro{BSs}{Base Stations}
\acro{BW}{bandwidth}
\acro{C-RAN}{Cloud Radio Access Networks}
\acro{CAPEX}{Capital Expenditure}
\acro{CoMP}{Coordinated Multipoint}
\acro{CR}{Cognitive Radio}
\acro{D2D}{Device-to-Device}
\acro{DAC}{Digital-to-Analog Converter}
\acro{DAS}{Distributed Antenna Systems}
\acro{DBA}{Dynamic Bandwidth Allocation}
\acro{DC}{Duty Cycle}
\acro{DL}{Deep Learning}
\acro{DSA}{Dynamic Spectrum Access}
\acro{FBMC}{Filterbank Multicarrier}
\acro{FEC}{Forward Error Correction}
\acro{FFR}{Fractional Frequency Reuse}
\acro{FSO}{Free Space Optics}
\acro{GA}{Genetic Algorithms}
\acro{HAP}{High Altitude Platform}
\acro{HL}{Higher Layer}
\acro{HARQ}{Hybrid-Automatic Repeat Request}
\acro{HCA}{Hierarchical Cluster Analysis}
\acro{HO}{Handover}
\acro{KNN}{k-nearest neighbors} 
\acro{IoT}{Internet of Things}
\acro{LAN}{Local Area Network}
\acro{LAP}{Low Altitude Platform}
\acro{LL}{Lower Layer}
\acro{LoS}{Line of Sight}
\acro{LTE}{Long Term Evolution}
\acro{LTE-A}{Long Term Evolution Advanced}
\acro{MAC}{Medium Access Control}
\acro{MAP}{Medium Altitude Platform}
\acro{MDP}{Markov Decision Process}
\acro{ML}{Machine Learning}
\acro{MME}{Mobility Management Entity}
\acro{mmWave}{millimeter Wave}
\acro{MIMO}{Multiple Input Multiple Output}
\acro{NFP}{Network Flying Platform}
\acro{NFPs}{Network Flying Platforms}
\acro{NLoS}{Non-Line of Sight}
\acro{OFDM}{Orthogonal Frequency Division Multiplexing}
\acro{O-RAN}{Open Radio Access Network}
\acro{OSA}{Opportunistic Spectrum Access}
\acro{PAM}{Pulse Amplitude Modulation}
\acro{PAPR}{Peak-to-Average Power Ratio}
\acro{PGW}{Packet Gateway}
\acro{PHY}{physical layer}
\acro{PSO}{Particle Swarm Optimization}
\acro{PU}{Primary User}
\acro{QAM}{Quadrature Amplitude Modulation}
\acro{QoE}{Quality of Experience}
\acro{QoS}{Quality of Service}
\acro{QPSK}{Quadrature Phase Shift Keying}
\acro{RF}{Radio Frequency}
\acro{RL}{Reinforcement Learning}
\acro{RMSE}{Root Mean Squared Error}
\acro{RN}{Remote Node}
\acro{RRH}{Remote Radio Head}
\acro{RRC}{Radio Resource Control}
\acro{RRU}{Remote Radio Unit}
\acro{RSS}{Received Signal Strength}
\acro{SU}{Secondary User}
\acro{SCBS}{Small Cell Base Station}
\acro{SDN}{Software Defined Network}
\acro{SNR}{Signal-to-Noise Ratio}
\acro{SON}{Self-organising Network}
\acro{SVM}{Support Vector Machine}
\acro{TDD}{Time Division Duplex}
\acro{TD-LTE}{Time Division LTE}
\acro{TDM}{Time Division Multiplexing}
\acro{TDMA}{Time Division Multiple Access}
\acro{UE}{User Equipment}
\acro{UAV}{Unmanned Aerial Vehicle}
\acro{USRP}{Universal Software Radio Platform}
\acro{DRL}{Deep Reinforcement Learning}
\acro{AI}{Artificial Intelligence}
\acro{RAN}{Radio Access Network}
\acro{RU}{Radio Unit}
\acro{CU}{Central Unit}
\acro{DU}{Distributed Unit}
\acro{NR}{New Radio}
\acro{gNBs}{Next Generation Node Bases}
\acro{CP}{Control Plane}
\acro{UP}{User Plane}
\acro{FPGAs}{Field  Programmable  Gate  Arrays}
\acro{ASICs}{Application-specific Integrated Circuits}
\acro{PHY-low}{lower level PHY layer processing}
\acro{FFT}{Fast Fourier Transform}
\acro{RRC}{Radio Resource Control}
\acro{SDAP}{Service Data Adaptation Protocol}
\acro{PDCP}{Packet Data Convergence Protocol}
\acro{RLC}{Radio Link Control}
\acro{RIC}{RAN Intelligent Controller}
\acro{RICs}{RAN Intelligent Controllers}
\acro{KPMs}{Key Performence Measurements}
\acro{RT}{Real Time}
\acro{SMO}{Service Management and Orchestration}
\acro{UE}{User Equipment}
\acro{API}{Application Programming Interface}
\acro{OSC}{O-RAN Software Community}
\acro{DRL}{Deep Reinforcement Learning}
\acro{OSP}{Online Service Provider}
\acro{NIB}{Network Information Base}
\acro{SDL}{Shared Data Layer}
\acro{SLA}{Service Level Agreement}
\acro{A1AP}{A1 Application Protocol}
\acro{HTTP}{Hypertext Transfer Protocol}
\acro{SL}{Supervised Learning}
\acro{UL}{Unsupervised Learning}
\acro{RL}{Reinforcement Learning}
\acro{DL}{Deep Learning}
\acro{FDD}{Frequency-division Duple}
\acro{TDD}{Time-division Duple}
\acro{LSTM}{Long Short-term Memory}
\acro{PCA}{Principal Component Analysis}
\acro{ICA}{Independent Component Analysis}
\acro{MDP}{Markov Decision Process}
\acro{GRL}{Generalization Representation Learning}
\acro{SRL}{Specialization Representation Learning}
\acro{SVM}{Support Vector Machine}
\acro{TDNN}{Time-delay Neural Network}
\acro{LSTM}{Long Short-term Memory}
\acro{MSE}{Mean Squared Error}
\acro{CNN}{Conventional Neural Network}
\acro{NAS}{Neural Architecture Search}
\acro{SDS}{Software Defined Security}
\acro{SON}{Self-organized Network}
\acro{KPIs}{Key Performance Indicators}
\acro{HetNet}{Heterogeneous Network}
\acro{HPN}{High Power Node}
\acro{LPNs}{Low Power Nodes}
\acro{QL}{Q-Learning}
\acro{PG}{Policy Gradient}
\acro{A2C}{Actor-Critic}
\acro{TD}{Temporal Defence}
\acro{SHAP}{SHapley Additive exPlanations}
\acro{DNNs}{Deep Neural Networks}
\acro{DNN}{Deep Neural Network}
\acro{MLP}{Multiple-layer Perceptron}
\acro{RNN}{Recurrent Neural Network}
\acro{DRL}{Deep Reinforcement Learning}
\acro{DQN}{Deep Q-Learning Network}
\acro{DDQN}{Double Deep Q-Learning Network}
\acro{GNN}{Graph Neural Network}
\acro{eMMB}{enhanced mobile broadband}
\acro{URLLC}{ultra-reliable low-latency communication}
\acro{QoS}{Quality of Service}
\acro{ILP}{Integer Linear Programming}
\acro{NF}{Network Function}
\acro{VFN}{Network Function Virtualization}
\acro{NBC}{Navie Bayes Classifier}
\acro{RAT}{Radio Access Technology}
\acro{FML}{federated meta-learning}
\acro{TS}{Traffic Steering}
\acro{CQL}{Conservative Q-Learning}
\acro{REM}{Random Ensemble Mixture}
\acro{SCA}{Successive Convex Approximation}
\acro{XAI}{eXplainable Artificial Intelligent}
\acro{D-RAN}{Distributed RAN}
\acro{C-RAN}{Cloud RAN}
\acro{v-RAN}{Virtual RAN}
\acro{RRH}{Remote Radio Head}
\acro{mMTC}{massive machine-type communication}
\acro{CDMA}{Code Division Multiple Access}
\acro{TDMA}{Time Division Multiple Access}
\acro{OFDMA}{Orthogonal Frequency-Division Multiple Access}
\acro{RRM}{Radio Resources Management}
\acro{NFV}{Network Function Virtualization}
\acro{D-RAN}{Distributed-RAN}
\acro{V-RAN}{Vritualized-RAN}
\acro{PA}{Power Amplifier}
\acro{SINR}{Signal to interference plus noise ratio}
\acro{mmWave}{millimeter Wave }
\acro{LOS}{Line of Sight}
\acro{NLOS}{Non-Line of Sight}
\acro{FSPL}{Free Space Path Loss}
\acro{EEMP}{Energy Efficiency Maximization Problem}
\acro{QFMEE}{QoS First Maximum EE}
\acro{SWES}{Switching-on/off based Energy Saving}
\acro{APC}{Area Power Consumption}
\acro{EE}{Energy Efficiency}
\acro{Near-RT}{Near-Real Time}
\acro{Open-RAN}{Open Radio Access Network}
\acro{Near-RT RIC}{Near Real Time RIC}
\acro{Non-RT RIC}{Non Real Time RIC}
\acro{TD3}{Twin Delayed Deep Deterministic Policy Gradient}
\acro{RC}{Radio Card}
\acro{UMa}{Urban Microcell path loss}
\acro{PRB}{Physical Resource Block}
\acro{API}{Application Programming Interface}
\acro{RSS}{Received Signal Strength}
\acro{DL}{downlink}
\acro{UL}{uplink}
\acro{DDPG}{Deep Deterministic Policy Gradient}
\acro{TD}{Temporal Difference}
\acro{RSRP}{Reference Signals Received Power}
\acro{FL}{Federated Learning}
\acro{UMi}{Urban Microcell}
\acro{BN}{Batch Normalization}
\acro{SBS}{Small Base Station}
\acro{FRL}{Federated Reinforcement Learning}
\acro{DQNSA}{DQN-Single Action}
\acro{DQNMA}{DQN-Multiple Action}
\end{acronym}

\bibliographystyle{IEEEtran}

\bibliography{References.bib}
\end{document}